# The coherent artifact in modern pulse measurements

Justin Ratner,[1] Günter Steinmeyer,[2] Tsz Chun Wong,[1*] Randy Bartels,[3] and Rick Trebino[1]

[1]*Georgia Institute of Technology, School of Physics, 837 State St NW, Atlanta, GA 30332 USA*
[2]*Max Born Institute for Nonlinear Optics and Short Pulse Spectroscopy, 12489 Berlin, Germany*
[3]*Colorado State University, Department of Electrical and Computer Engineering, Fort Collins, CO USA*
*jeff.wong@gatech.edu

We simulate multi-shot intensity-and-phase measurements of unstable ultrashort-pulse trains using frequency-resolved-optical-gating (FROG) and spectral phase interferometry for direct electric-field reconstruction (SPIDER). Both techniques fail to reveal the pulse structure. FROG yields the average pulse duration and suggests the instability by exhibiting disagreement between measured and retrieved traces. SPIDER under-estimates the average pulse duration but retrieves the correct average pulse spectral phase. An analytical calculation confirms this behavior.

When a measurement averages over many different events, it faces an impossible task: providing one result when no single result can be correct. In ultrafast optics, this issue has been particularly problematic for multi-shot intensity-autocorrelation measurements of trains of different complex pulses. The resulting measured autocorrelation trace vs. delay (see Fig. 1) consists of a narrow spike atop a broad structureless background.[1] Although early researchers often mistook the spike, or *coherent artifact*, for the measure of their pulse length, its width actually only yields the much shorter *nonrandom* component of the pulse. The correct pulse length is actually indicated by the temporally much broader background, which also takes into account the much longer, *random*, or incoherent, pulse component.

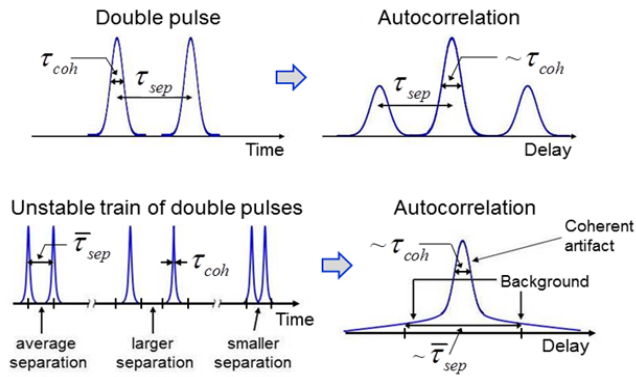

Fig. 1. (Color online) Top: Double pulse and its autocorrelation. Bottom: A train of variably spaced double pulses and their multi-shot autocorrelation. The coherent artifact results from the short nonrandom coherent component of the double pulses (a single pulse), while the broader background results from the overall average pulse length (the combination of both pulses). This trace is typical of autocorrelations of nearly all trains of unstable complex pulses.

Given that the task is inherently impossible, it is worth asking what we should expect. The answer is that the technique should yield a pulse with some characteristics of the typical pulse in the train (e.g., its duration) and also give some indication of the stability, or randomness, of the pulses in the train. Although autocorrelation actually does yield some of this information, it yields neither the pulse intensity nor its phase for the case of a *stable* train of identical pulses and so is now generally considered obsolete.

The next question—one whose answer is long overdue—is how more modern pulse-measurement techniques, which do yield the pulse intensity and phase for a stable train of identical pulses, react to an unstable train of random pulses. So here we consider this question for frequency-resolved optical gating (FROG)[2] and spectral-phase interferometry for direct electric-field reconstruction (SPIDER)[3], the latter of which also allows an analytical result.

For the simulations, we chose a nonrandom component $E(\omega)$ with a flat phase and Gaussian intensity of temporal FWHM $12\delta t$, where $\delta t$ is the temporal sampling rate. To $E(\omega)$, we added an equal-energy random component $E_{rand}(\omega)$ with the same spectrum, but with random spectral phase, which we then Fourier-filtered (made somewhat less random and hence the resulting pulse shorter) by different amounts to yield two trains of variably structured pulses with different average complexities and durations.[4] The random trains' resulting average pulse lengths (FWHM) were $26\delta t$ and $54\delta t$. Figure 2 shows typical pulses in the two trains. All frequency units are in $2\pi/(N\delta t)$, where $N$ is the SPIDER array size (4096).

We computed multi-shot traces for second-harmonic-generation (SHG) FROG and SPIDER. FROG involves measuring a self-gated spectrogram of the pulse field, while SPIDER measures a spectral interferogram of the pulse and a frequency-sheared and delayed replica of it. SPIDER requires a frequency shear, $\delta\omega$, which we chose to be $30\pi/(N\delta t)$, and a pulse separation, $T$, which we chose to be $300\delta t$. For both techniques, our simulated traces averaged over the same trains of 5000 different pulses. Finally, we retrieved pulses from the traces using the generalized projections SHG FROG algorithm[2] and the Takeda algorithm[5] for SPIDER (see Fig. 2).

Note the narrow autocorrelation-like coherent artifacts in the centers of the random-train FROG traces. Despite this, FROG retrieves the correct approximate average pulse durations: 27 and 52$\delta t$. However, the retrieved pulses are simple, lacking the structure of the actual pulses in the unstable trains.

In addition, note the "retrieved" FROG traces, which provide a consistency check on the measurement. This is because FROG traces (two-dimensional arrays) have many more points than do pulses (pairs of one-dimensional arrays), so any systematic error in a measured trace preventing the algorithm from retrieving the correct pulse is revealed by disagreement between the measured and retrieved traces. Indeed, because the multi-shot FROG trace for each unstable pulse train does not correspond to any single pulse, there is significant disagreement and a high rms difference ("G error") between the two traces, indicating that something is amiss. Of course, it is up to the user to determine precisely what.

For all three pulse trains, we find that SPIDER accurately yields the flat spectral phase of the *nonrandom* component, and hence a pulse 12$\delta t$ long, independent of the random component. Thus, SPIDER *under*-estimates the average durations by factors of 2.2 and 4.5. Instability is indicated by <100% SPIDER fringe visibilities $V$ of 98% and 90%, respectively. Reduced SPIDER fringe visibility is usually ignored because it can also arise from benign misalignment effects. The black dotted SPIDER traces in Fig. 2 are fits to the unstable-pulse-train traces, assuming instead a stable flat-phase Gaussian pulse and one such benign effect: unequal energies of the SPIDER-device pulse pair. Note that these fits are indistinguishable from the traces for the unstable pulse trains.

SPIDER allows a simple calculation of its trace in the presence of a random component, which confirms these results. The multi-shot SPIDER trace is:

$$S_{SPIDER} \propto \left\langle \left| E(\omega) + E(\omega + \delta\omega)\exp(i\omega T) + E_{rand}(\omega) + E_{rand}(\omega + \delta\omega)\exp(i\omega T) \right|^2 \right\rangle \quad (1)$$

where the brackets indicate a multi-pulse average. Expanding this expression, noting that terms containing only one random field sum to zero in the average (due, for example, to simple zero[th]-order spectral-phase variations), and writing in terms of the spectra, $S(\omega)$ and $S_{rand}(\omega)$, spectral phases, $\varphi(\omega)$ and $\varphi_{rand}(\omega)$, and group delays vs. frequency, $\tau(\omega) = d\varphi/d\omega$ and $\tau_{rand}(\omega) = d\varphi_{rand}/d\omega$, for the two pulse components:

$$S_{SPIDER} = S(\omega) + S(\omega + \delta\omega) + \langle S_{rand}(\omega)\rangle + \langle S_{rand}(\omega + \delta\omega)\rangle + 2\sqrt{S(\omega)}\sqrt{S(\omega + \delta\omega)}\cos[\delta\omega\ \tau(\omega) + \omega T] + 2\left\langle \sqrt{S_{rand}(\omega)}\sqrt{S_{rand}(\omega + \delta\omega)}\cos[\delta\omega\ \tau_{rand}(\omega) + \omega T]\right\rangle \quad (2)$$

The first line is the sum of the spectra. The second is the usual SPIDER fringe term from which the pulse spectral phase is retrieved, but only for the *nonrandom* component of the pulse. The last is the SPIDER fringe term for the random component of the pulse.

Clearly, variations in the spectral phase of the random component, even just the pulse arrival time $\tau_{rand}(\omega_0)$ (the first-order spectral-phase component), will cause the last term to wash out, leaving only the spectral background and the SPIDER term for the coherent nonrandom component of the pulse. Specifically, arrival-time variations over a range of $2\pi/\delta\omega$ clearly wash this term out completely, rendering a variable-delay satellite pulse invisible to SPIDER (the case of Fig. 1), an effect anticipated in [6].

As this arrival-time effect is clear from Eq. (2), we removed it in our simulations by centering the random component on the nonrandom one. Still, cancellations occur for *any* randomness in the spectral phase and have done so in our simulations, despite our additional spectral-phase smoothing to reduce the complexity and duration of the random components. However, there will be extra background of $\langle S_{rand}(\omega)\rangle + \langle S_{rand}(\omega + \delta\omega)\rangle$, which will reduce the fringe visibility $V$. Alas, the measurement will say little about the random component's spectral phase.

To summarize, we find that SPIDER retrieves an excellent estimate of the *average* spectral phase and the *nonrandom* component of the pulse train. But it does *not* see any randomly varying component of the pulse. In short, for an unstable pulse train, multi-shot SPIDER measures only the *coherent artifact*.

This should not be surprising: interferometric methods, in general, are not sensitive to random phase variations, responding only with reduced fringe visibility and increased background. We are unaware of any SPIDER measurements with fringe visibilities greater than 98%, a value that, in our simulations, corresponds to a measured pulse length too short by more than a factor of 2. Indeed, in supercontinuum measurements, much smaller values—as low as 10%—have been reported[7]. Without deeper insight into the underlying physics[8] or additional independent measurements, it appears impossible to determine whether an imperfect SPIDER fringe visibility is due to benign alignment effects (and so corresponds to a stable train of short pulses) or instability (and so corresponds to an unstable train of potentially much longer ones). Thus, unless the pulse-to-pulse stability of the temporal intensity can otherwise be ensured, it appears that pulse-length claims from measurements with imperfect SPIDER fringe visibility require re-evaluation.

Some FROG measurements will also require re-evaluation. While SHG FROG yields the correct pulse lengths in our simulations, it, like SPIDER, misses the pulse structure and so could also yield misleading results in the presence of instability. However, FROG provides a strong indicator of instability: disagreement between the measured and retrieved FROG traces. Unfortunately, some authors have attributed such disagreement to possible nonconvergence of the FROG algorithm. In view of our results and the FROG algorithm's demonstrated robustness for all but extremely complex pulses,[4] such discrepancies appear much more likely to be due to instability. Fortunately, instability is, in fact, more often considered as the cause, having previously been encountered experimentally in cross-correlation FROG (XFROG) measurements of supercontinuum pulse trains[9]. In that case, XFROG retrieved a pulse with the extreme complexity of a typical pulse in the train, and it was the disagreement between the measured and retrieved traces that indicated a problem and inspired single-shot spectral measurements and extensive theoretical investigations,[10] confirming the highly unstable nature of the continuum.

In any case, whichever technique is used, measured—and, if available, retrieved—traces should always be reported. Only good agreement between measured and retrieved FROG traces or a 100% SPIDER fringe visibility can imply good pulse-train stability.

The authors acknowledge support from the National Science Foundation, Grant #ECCS-1028825 and the Georgia Research Alliance.

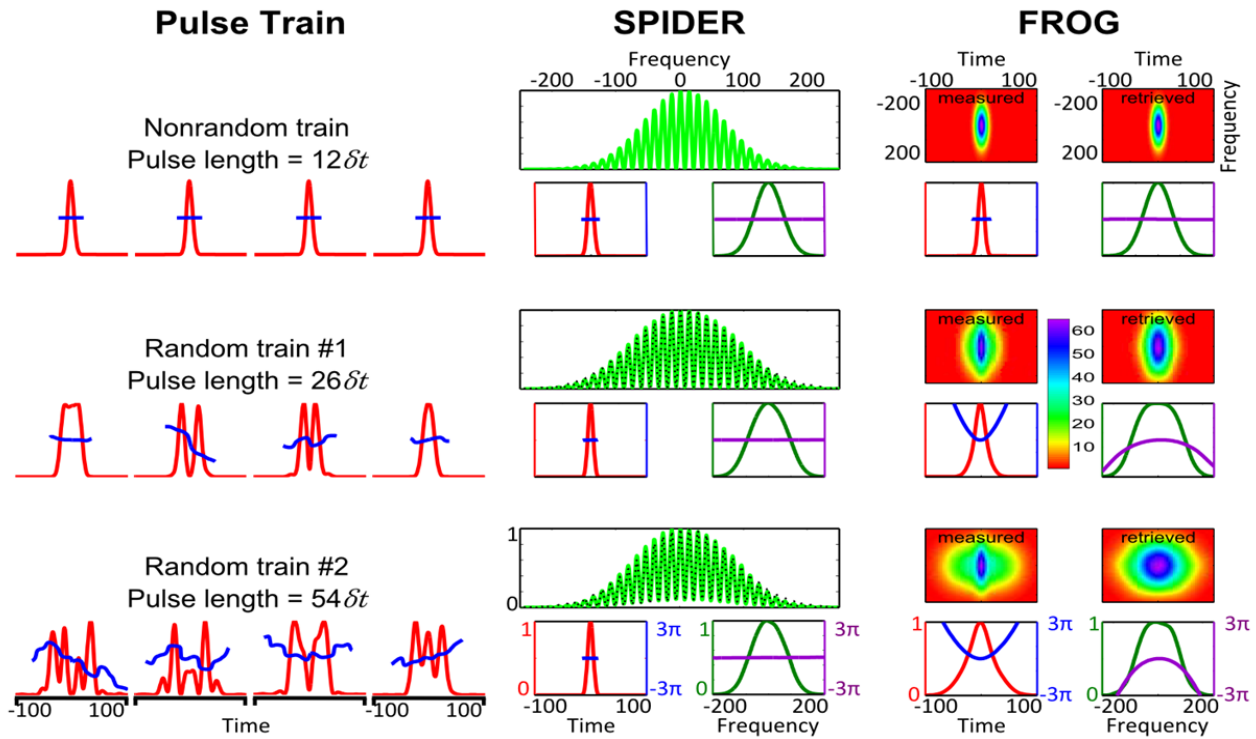

Fig. 2. (Color online) Nonrandom- and random-pulse trains of varying complexity, and simulated multi-shot SPIDER and SHG FROG measurements of them. Top row: nonrandom train of identical Gaussian flat-phase pulses. Middle and bottom rows: random-pulse trains of different average complexity and duration. Red curves indicate intensity, blue phase, green spectrum, and purple spectral phase. The black dotted SPIDER traces are fits assuming flat-phase Gaussian pulses and unequal SPIDER double-pulse energies. For all three pulse trains, SPIDER retrieves only the nonrandom pulse component, $12\delta t$ long, and exhibits decreasing fringe visibility (100%, 98%, and 90%, respectively). FROG also does not see the pulse structure but does yield the correct durations. Measured and retrieved FROG traces also disagree for the random trains, and their rms differences (G errors) are large: 0.0083 and 0.014, respectively, for the 256×256 traces. Both techniques retrieve the stable train (top row) perfectly.

**References**

1. E. P. Ippen, and C. V. Shank, *Ultrashort Light Pulses—Picosecond Techniques and Applications* (Springer-Verlag, 1977).
2. R. Trebino, *Frequency-Resolved Optical Gating: The Measurement of Ultrashort Laser Pulses* (Kluwer Academic Publishers, 2002).
3. C. Iaconis, and I. A. Walmsley, Opt. Lett. **23**, 792-794 (1998).
4. L. Xu, E. Zeek, and R. Trebino, J. Opt. Soc. Am. B **25**, A70-A80 (2008).
5. M. Takeda, H. Ina, and S. Kobayashi, J. Opt. Soc. Am. **72**, 156-160 (1982).
6. F. Quéré, J. Itatani, G. Yudin, and P. Corkum, Phys. Rev. Lett. **90** (2003).
7. M. Yamashita, M. Hirasawa, N. Nakagawa, K. Yamamoto, K. Oka, R. Morita, and A. Suguro, J. Opt. Soc. Am. B **21**, 458-462 (2004).
8. K. L. Corwin, N. R. Newbury, J. M. Dudley, S. Coen, S. A. Diddams, K. Weber, and R. S. Windeler, Phys. Rev. Lett. **90**, 113904 (2003).
9. X. Gu, L. Xu, M. Kimmel, E. Zeek, P. O'Shea, A. P. Shreenath, R. Trebino, and R. S. Windeler, Opt. Lett. **27**, 1174-1176 (2002).
10. X. Gu, M. Kimmel, A. P. Shreenath, R. Trebino, J. Dudley, S. Coen, and R. S. Windeler, Opt. Expr. **11**, 2697-2703 (2003).